\documentclass[]{llncs}

\usepackage[T1]{fontenc}
\usepackage{lmodern}
\usepackage{amssymb,amsmath}
\usepackage{ifxetex,ifluatex}
\usepackage{csvsimple}
\usepackage{longtable,booktabs}
\usepackage{makeidx}
\institute{Support Centre for Advanced Neuroimaging, University Institute of Diagnostic and Interventional Neuroradiology, Inselspital, Bern University Hospital, Bern, Switzerland  \\ \texttt{}}

\usepackage[math]{cellspace}
\IfFileExists{upquote.sty}{\usepackage{upquote}}{}
\ifnum 0\ifxetex 1\fi\ifluatex 1\fi=0 
  \usepackage[utf8]{inputenc}
\else 
  \ifxetex
    \usepackage{mathspec}
    \usepackage{xltxtra,xunicode}
  \else
    \usepackage{fontspec}
  \fi
  \defaultfontfeatures{Mapping=tex-text,Scale=MatchLowercase}
  
\fi
\IfFileExists{microtype.sty}{\usepackage{microtype}}{}
\usepackage{graphicx}
\makeatletter
\def\ScaleIfNeeded{%
  \ifdim\Gin@nat@width>\linewidth
    \linewidth
  \else
    \Gin@nat@width
  \fi
}
\makeatother

\ifxetex
  \usepackage[setpagesize=false, 
              unicode=false, 
              xetex]{hyperref}
\else
  \usepackage[unicode=true]{hyperref}
\fi

\title{Uncertainty-driven refinement of tumor-core segmentation using 3D-to-2D networks with label uncertainty.}
\author{Richard McKinley, Micheal Rebsamen, Katrin D\"atwyler, Raphael Meier, Piotr Radojewski, Roland Wiest}
\date{}

\begin{document}
\maketitle
\begin{abstract}
The BraTS dataset contains a mixture of high-grade and low-grade gliomas, which have a rather different appearance: previous studies have shown that performance can be improved by separated training on low-grade gliomas (LGGs) and high-grade gliomas (HGGs), but in practice this information is not available at test time to decide which model to use.  By contrast with HGGs, LGGs often present no sharp boundary between the tumor core and the surrounding edema, but rather a gradual reduction of tumor-cell density.

Utilizing our 3D-to-2D  fully convolutional architecture, DeepSCAN, which ranked highly in the 2019 BraTS challenge and was trained using an uncertainty-aware loss,  we separate cases into those with a confidently segmented core, and those with a vaguely segmented or missing core.  Since by assumption every tumor has a core, we reduce the threshold for classification of core tissue in those cases where the core, as segmented by the classifier,  is vaguely defined or missing. 

We then predict survival of high-grade glioma patients using a fusion of linear regression and random forest classification, based on age, number of distinct tumor components, and number of distinct tumor cores.

We present results on the validation dataset of the Multimodal Brain Tumor Segmentation Challenge 2020 (segmentation and uncertainty challenge), and on the testing set, where the method achieved 4th place in Segmentation, 1st place in uncertainty estimation, and 1st place in Survival prediction.


\end{abstract}

\section{Introduction}

The BRATS challenge \cite{Menze2015,Bakas2018}, and its accompanying dataset \cite{Bakas2017,Bakas2017a,Bakas2017b} of
annotated glioma images have driven substantial amounts of research in brain tumor segmentation. \cite{Havaei2017BrainTS,Kamnitsas2017EfficientM3,Perreira2016}  The data is taken from various sources/centers/scanners and incorporates both low-grade gliomas (LGGs) and high-grade gliomas (HGGs).  While this may enhance the applicability of models trained on the dataset, previous work has shown that the inclusion of both low- and high-grade tumors (which have rather different visual appearance) leads to lower performance than training on only low- or high-grade tumors \cite{Rebsamen2019}.

Based on our 3rd-place entry to the BRaTS 2018/2019 segmentation challenge  \cite{Mckinley2018,mckinley2020}, which was trained using an uncertainty-aware loss function which outputs confidence maps for each voxel/tissue type, we attempt to solve the problem of uncertain core segmentation in LGGs.  We observe that in the subclass of LGGs which have poorly a delineated core, the whole core is relatively uncertain, whereas in cases with a well-segmented core, the center of the core is predicted with very high certainty.  We hypothesize that this arises in diffuse LGGs that have practically no visible boundary between the solid tumor core and the surrounding edema. This also reflects tumor biology: the cell density of tumor cells does not abruptly fall to zero, but rather gradually reduces.  We attempt to compensate for this effect by reducing the threshold at which voxels are classified as tumor core, in cases where the core is poorly delineated (as defined by the confidence of the core segmentation).  This is essentially the opposite approach to Nair et al. \cite{Nair2018,nair2020a}, where lesions with high uncertainty are deleted from the lesion mask.

\section{Heteroscedastic label-flip loss and focal KL-divergence}
In our previous BraTS submission\cite{mckinley2020}, for the training of our model we used a \emph{label-flip} loss: for each voxel and each tissue type, the classifier produces an output $p \in (0,1)$ and an output $q \in (0,0.5)$ which represents the probability that the classifier output differs from the ground truth.\footnote{The range of values allowed for $q$ tends to cause confusion, with many readers expecting to find values in a range $(0,1)$, since $q$ is framed as a probability.  However, for a binary classification problem, an uncertainty of $0.5$ indicates that the classifier is, in effect, randomly guessing.  A $q$ greater than $0.5$ would correspond to a classifier which predicts e.g. label $1$ but believes that the label in the ground truth is $0$.}  The label flip loss function used in our previous Brats submission had the form:

\begin{equation}
\mathrm{Focal}(p, (1-x)*q + x*(1-q)) + \mathrm{BCE}(q, z)
\label{eq:fliploss}
\end{equation}

where 
z
is the indicator function for disagreement between the classifier (thresholded at the $p=0.5$ level) and the ground truth, BCE is binary cross-entropy, and Focal is \emph{focal loss}.~\cite{Lin2017FocalLF}

This can be understood as a form of heteroscedastic classification networks (a network which predicts the variance of their preactivation outputs) as introduced in \cite{Kendall2017WhatUD}.  The loss function has the advantage that (unlike predicting the variance of logit outputs as in Kendall et al) it is differentiable in $p$ and $q$ and so can be backpropogated through. The focal loss term aims to separate tissue from non-tissue voxels, with attention paid to those close to the decision boundary (focal loss attenuates gradient contributions from well-classified examples far from the decision boundary).  This allows learning to focus on difficult-to-classify cases, meaning that the effect of class imbalance between foreground and background is lessened.  However, voxels close to the tumor border are inherently uncertain and are typically not consistently labeled by human raters.  We want learning to concentrate on examples which are incorrectly classified, and which are not inherently uncertain (voxels with a high probability of deviation from the reference 'ground truth').  Focal loss is therefore attenuated according to the probability of a 'label-flip' between the 
output of the classifier and the ground truth.

In practice, this loss function rarely produces values of $q$ close to 0.5.  We believe this is because the loss function cannot actively encourage examples to remain (correctly) classified as fundamentally uncertain: the loss function always produces a gradient, however weak, towards the human-provided ground truth.    For targets $w$ in $[0,1]$, the minimum value of BCE is the entropy $-w \; log(w)$ of $w$.  Therefore, if $w$ is close to 0.5, the loss can only be lowered by lowering the uncertainty.  However, many examples will indeed have uncertainty close to 0.5, in particular along the boundaries between tissue types. 
Subtracting the entropy yields the Kullback-Leibler divergence, which does not provide any incentive to reduce the uncertainty measurement.

\[  \mathrm{KL}(w \parallel p) = w \; log(w) - w \; log(p) \] 

KL divergence is typically not used as a loss in classification problems, because the entropy term $w \; log(w)$ is fixed: however, in our setting $w$ can vary.  In this year's submission we amended the loss function to incorporate a form of \emph{focal Kullback Leibler divergence}, which encourages highly uncertain examples to remain close to the decision boundary:

\[  \mathrm{Focal_{KL}}(w \parallel p) = (p - w)^2 (w\; log(w) - w \; log(p)) \]

The new loss function combines focal KL-divergence with binary cross-entropy for estimation of label uncertainty:
\[ \mathrm{Focal_{KL}}(w \parallel p) +  \; \mathrm{BCE}(q,z)\]

where 
z
is the indicator function for disagreement between the classifier (thresholded at the $p=0.5$ level) and $w$ is $(1-x)*q + x*(1-q)$. In this loss function (unlike the previous) the first term goes to zero as $p$ tend to $w = (1-x)*q + x*(1-q)$.  With this loss function we observe uncertainty values close to $0.5$, representing total uncertainty (essentially uncertain examples which should not be further learned from).

In practical experiments we found that a combination of focal loss and the label-flip loss performed best:  our final loss function was of the form:

\[  \lambda \; \mathrm{Focal}(p,x) + (1 - \lambda) \; \mathrm{Focal_{KL}}(w \parallel p) + (1 - \lambda) \; \mathrm{BCE}(q,z)\]

with $\lambda = 0.1$ giving good results for retraining models already trained with the previous loss function.

\subsection{Ensembling and label uncertainty}
To ensemble the output of several models, we need to combine the output of the model into a single score.  Simply taking the mean of the $p_i$ and the mean of the $q_i$ does not adequately reflect the joint opinion of the models.  To see this, observe that for ensembling two predictions, if $p_1=0, q_1 = 0$ and  $p_2 =1, q_1 = 0$, the plain ensemble (by averaging) will yield a prediction of $0.5$ with probability $0$ of disagreeing with the human ground truth.  For a single model output, the function $f(p,q) = (1-x)*I_{p<=0.5} + x*(1-q) I_{p>=0.5}$ gives a value between 0 and 1 which combines the prediction and the uncertainty of the model and can be ensembled with other predictions in an ordinary fashion: given $n$ predictions $p_1 \dots p_n$ and $q_1 \dots q_n$,  the mean of $f(p_i, q_i)$ provides a single ensembled prediction.

 \section{Application to Brain Tumor Segmentation}
 
 \subsection{Data preparation and homogenization}

The raw values of MRI sequences cannot be compared across scanners and
sequences, and therefore a homogenization is necessary across the
training examples. In addition, learning in CNNs proceeds best when the
inputs are standardized (i.e.~mean zero, and unit variance). To this
end, the nonzero intensities in the training, validation and testing
sets were standardized, this being done across individual volumes rather
than across the training set. This achieves both standardization and
homogenization. 

\subsection{The DeepSCAN architecture with attention}
                \begin{figure}
                \centering
                	\vspace*{0.8cm}                    \includegraphics[width=\linewidth]{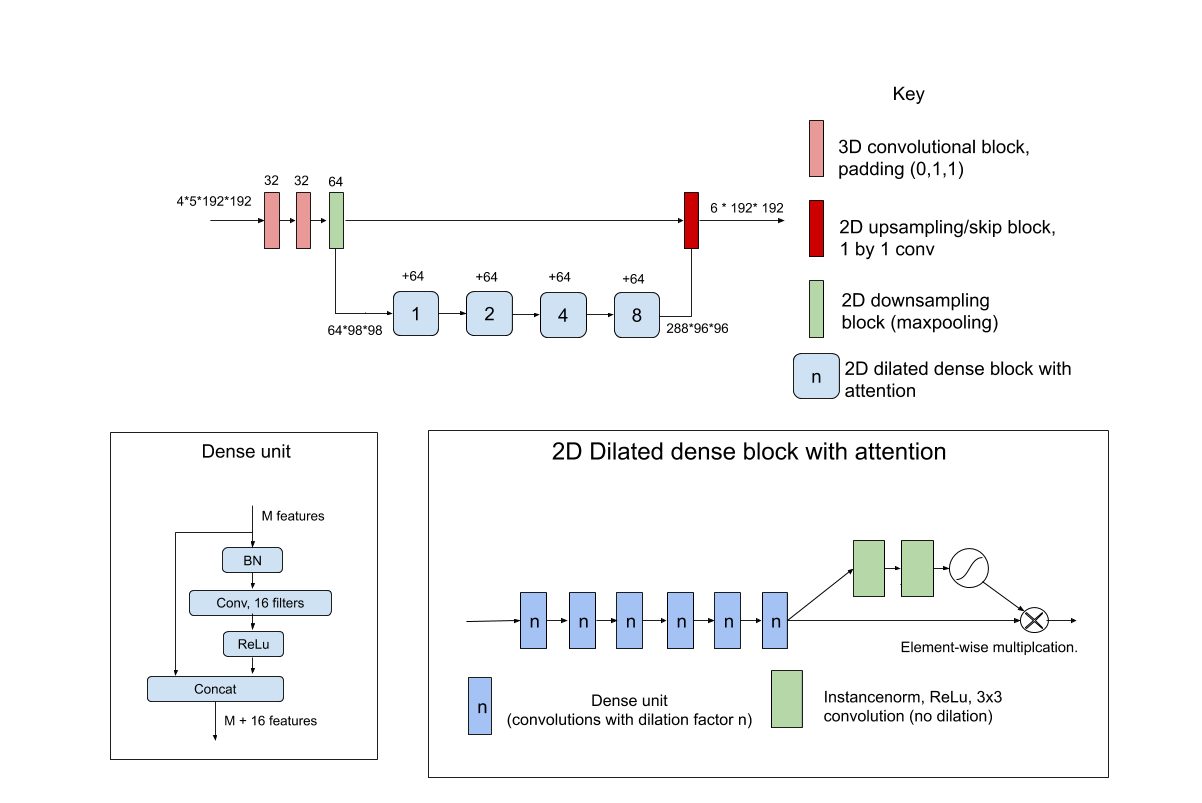}
                    
                    \caption{The DeepSCAN classifier, as applied in this paper to Brain Tumor Segmentation}
                    \label{fig:deepscan}
				\end{figure}
Our model architecture (shown in Figure \ref{fig:deepscan}) is identical to our entry to the 2019 challenge: this model has also been applied to the segmentation of MS lesions \cite{McKinley2019SimultaneousLA}, segmentation of brain anatomy \cite{McKinley2019c}, and measurement of cortical thickness \cite{rebsamen}.  The network was implemented in Pytorch: it consists of an initial phase of 3D convolutions to reduce a non-isotropic 3D patch to 2D, followed by a shallow encoder/decoder network using densely connected dilated convolutions in the bottleneck. This architecture is very similar to that used in our BraTS 2018 submission: principal differences are that we use Instance normalization rather than Batch normalization, and that we add a simple local attention mechanism of our own design (which element-wise multiplies the dilated feature maps with a mask calculated using non-dilated convolutions followed by a sigmoid) between dilated dense blocks.  In testing in 2019 we found that adding this attention mechanism was more effective than adding additional dense block, or making dense blocks wider.

We use multi-task rather than multi-class classification: each tumor region (Whole tumor, tumor core, enhancing tumor) is treated as a separate binary classification problem.   Inputs to the network (5*196*196 patches) were sampled randomly from either axial, sagittal or coronal direction.  We perform simple data augmentation: reflection about the (approximate) midline, rotation around a random principal axis through a random angle, and global shifting/rescaling of voxel intensities.  The network was trained with ADAM, using a batch size of 2 and weight decay $10^{-5}$. Models were trained using five-fold cross-validation: to reduce the training time, we used the pre-trained models from our 2019 challenge entry  as initial weights, ensuing that the same cases to train each model as in previous training rounds (new training examples were distributed evenly across the folds). We employed  cosine annealing learning rate schedule with restarts~\cite{Loshchilov2016}, restarting every 20000 gradient steps (we refer to this rather loosely as an epoch). We trained our model for 10 epochs with learning rate annealing from $10^{-4}$ to $10^{-7}$, and then for 10 further epochs with learning rate annealing from $10^{-5}$ to $10^{-8}$.

Final segmentations were derived by ensembling (as described above) the axial, sagittal and coronal views from the five models obtained from 5-fold cross-validation, with test-time augmentation provided by flipping in the saggital direction and rotating input images through $45^\circ$ in each plane.

\section{Uncertainty-based filtering}

In common with many previous entries, we perform minimal filtering of the final segmentation (Small components ($<10$ voxels) of any tissue class were deleted).  In addition, we observed (in common with our 2019 contribution) that for some low-grade gliomas the tumor core was completely missed, or that small isolated sections of the solid tumor were correctly segmented, but the majority was missed.  Lowering the threshold for segmentation to 0.05 (from 0.5) meant that these solid tumors were correctly segmented, but led to a lot of false positive tumor core identification in high grade gliomas.  Since the challenge does not provide the grade of gliomas, we require a proxy for the grade: we observed that in tumors which were well segmented (mostly HGGs) the mean of the ensembled model output inside the segmented core tumor was above 0.9, whereas in tumors with a poorly segmented core (mostly LGGs) the mean of the ensembled model output inside the tumor (if any tissue was identified as core at all) was typically below 0.75. Thinking of the mean of the ensembled model output  as a confidence in the global performance of the model, we therefore set 0.75 as a mean tumor core confidence threshold, and in cases with mean core confidence below that threshold, we reset the threshold for core segmentation to 0.05, thus capturing more tissue as tumor core in those cases.  Similar filtering was also applied to whole tumor (confidence threshold of .90) and enhancing tumor (confidence threshold of 0.8).  If no tumor core was found after filtering, the whole tumor was set to be tumor core (as in our 2019 submission), on the biological basis that all gliomas should contain a solid core of tumor tissue. Finally, we have seen in our clinical cases that certain low-grade gliomas may be detected very poorly, with no voxels exceeding the 0.5 threshold for detection.  If no tumor was detected at all, the threshold for whole tumor detection was lowered until at least 1000 voxels of tumor were detected, on the assumption that the model is only applied to glioma cases.  This final 'failsafe' (which we did not observe being triggered in training or testing our model) means that the algorithm is only suitable for outlining  tumors in patients known to have a glioma (or similar tumor), not for automatically \emph{detecting} tumors.

\
\subsection{Results}
Results of our classifier, as applied to the official BraTS validation data from the 2020 challenge, as generated by the official BraTS validation tool, before and after filtering for uncertainty  are shown in Table \ref{val_results}.

\begin{table}
\centering
\begingroup
\setlength{\tabcolsep}{4pt} 
\renewcommand{\arraystretch}{1.5} 
\begin{tabular}{c|cccccc}
                & \multicolumn{3}{c|}{Dice}               & \multicolumn{3}{c}{Hausdorff 95} \\
                & ET   & WT   & \multicolumn{1}{c|}{TC}   & ET        & WT        & TC       \\ \hline
Raw output      & 0.76 & 0.90 & \multicolumn{1}{c|}{0.80} & 26.8      & 5.25      & 12.4     \\
Filtered Output & 0.76 & 0.91 & \multicolumn{1}{c|}{0.85} & 26.8      & 3.91      & 5.61    
\end{tabular}
\endgroup
\vspace{1em}
\caption{Results on the BRATS 2020 validation set using the online validation tool.  Raw output denotes the ensembled results of the five classifiers derived from cross-validation.}
\label{val_results}
\end{table}

\section{Uncertainty challenge}
The BraTS uncertainty challenge requires the submission of an uncertainty score (from 0 to 100) per voxel and tissue type, denoting the certainty of classification at that voxel from most certain (100) to least certain (0).  The uncertainty output of our network (denoted $q$ above) can be directly converted to such an uncertainty score by taking  \[100*(1 - 2q).\]  The uncertainty in our ensemble can likewise be extracted as for any ordinary model with a sigmoid output $x$ as 

\[ 100*(1- 2\lvert 0.5 - x \rvert ) \]

While this uncertainty measure gives a measure of uncertainty both inside and outside the provided segmentation, in practice the following uncertainty, which treats all positive predictions as certain, and only assigns uncertain values to negative predictions, performs better according to the challenge metrics (see Table \ref{uncert_results}):

\[ 200*\mathrm(max( 0.5 - x, 0))\]

\begin{table}
\centering
\begingroup
\setlength{\tabcolsep}{3pt} 
\renewcommand{\arraystretch}{1.5} 
\begin{tabular}{c|ccc|ccc|ccc}
                       & \multicolumn{3}{c|}{Dice AUC} & \multicolumn{3}{c|}{TP AUC} &   \multicolumn{3}{c}{TN AUC} \\
                       & WT   & TC       & ET   & WT    & TC     & ET    & WT    & TC     & ET    \\ \hline
Baseline               & 0.91 & 0.85     & 0.76 & 0     & 0      & 0     & 0     & 0      & 0     \\
All uncertainties      & 0.93 & 0.86     & 0.8  & 0.07  & 0.13   & 0.1   & 0.004 & 0.003  & 0.001 \\
Negative Uncertainties & 0.93 & 0.87     & 0.78 & 0 & 0  & 0 & 0.004 & 0.003  & 0.001

\end{tabular}
\endgroup
\vspace{1em}
\caption{Results on the BRATS 2020 validation set using the online validation tool: uncertainty challenge.}
\label{uncert_results}
\end{table}

\section{Survival Prediction}

The survival prediction challenge requires a prediction of overall survival (in days) for subjects with "Gross Tumor Resection": while the output should be a number of days, the validation is based on accuracy of classification into three classes: long-survivors (>15 months), short-survivors (<10 months), and mid-survivors (e.g., between 10 and 15 months).  The age of the cases is provided: all other predictors must be derived from the imaging.  Previous work has shown that patient age alone can predict patient outcome relatively well, and outperform approaches integrating more complicated radiomic features.~\cite{weninger2019,kofler2020}

\begin{figure}
                \centering
                	\vspace*{0.9cm}
                    \includegraphics[width=0.4\linewidth]{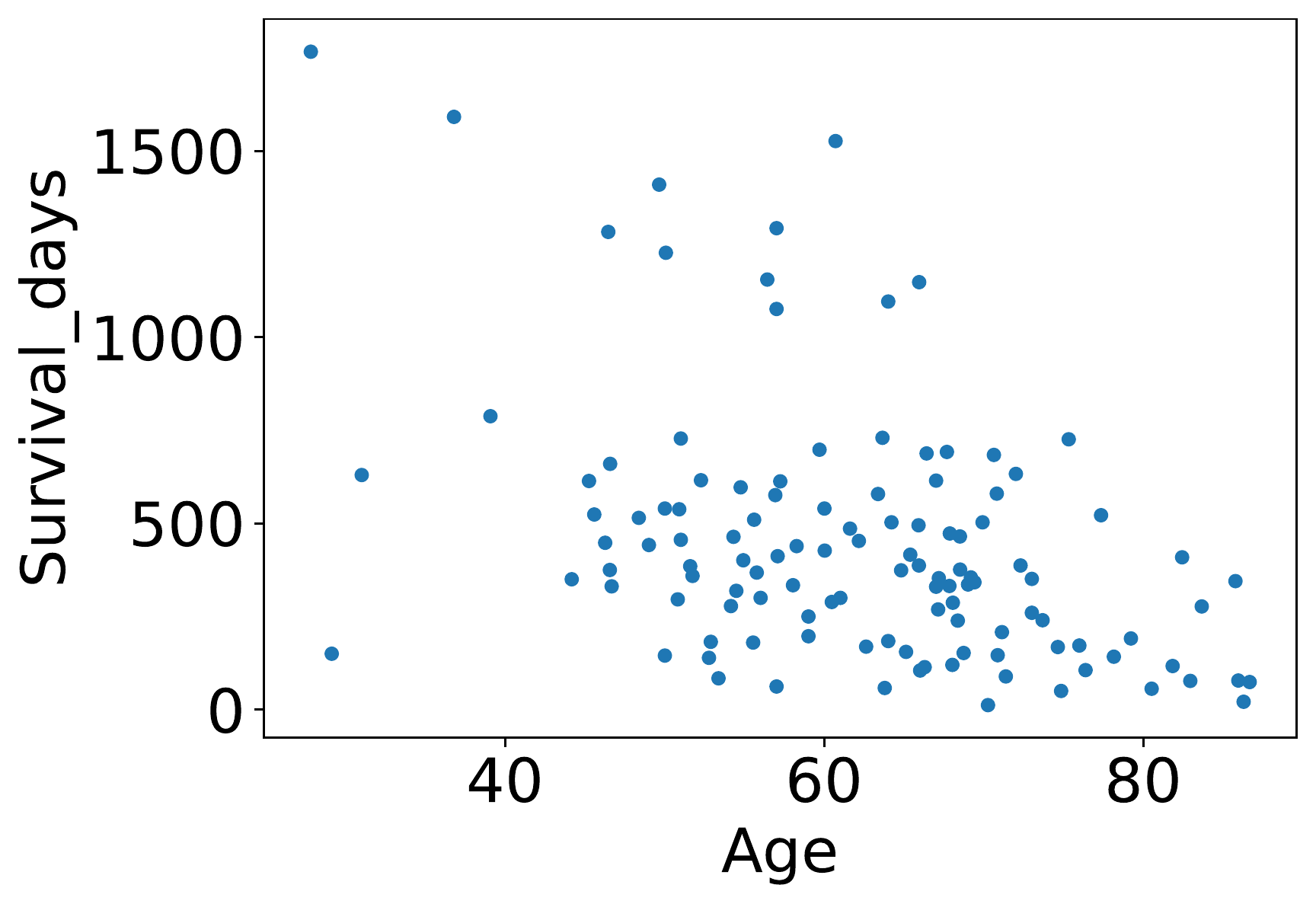}
                    \includegraphics[width=0.4\linewidth]{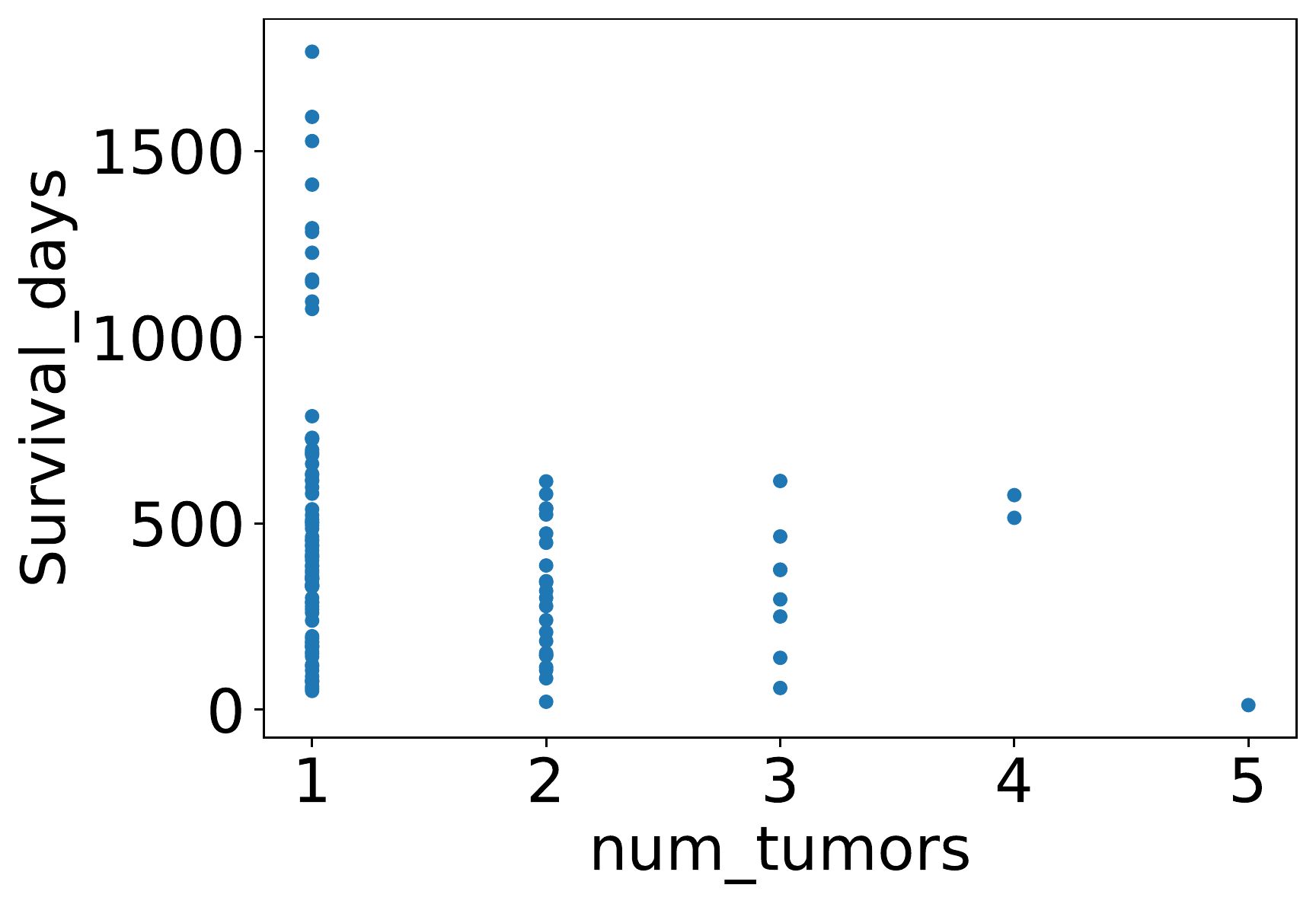}
                    \includegraphics[width=0.4\linewidth]{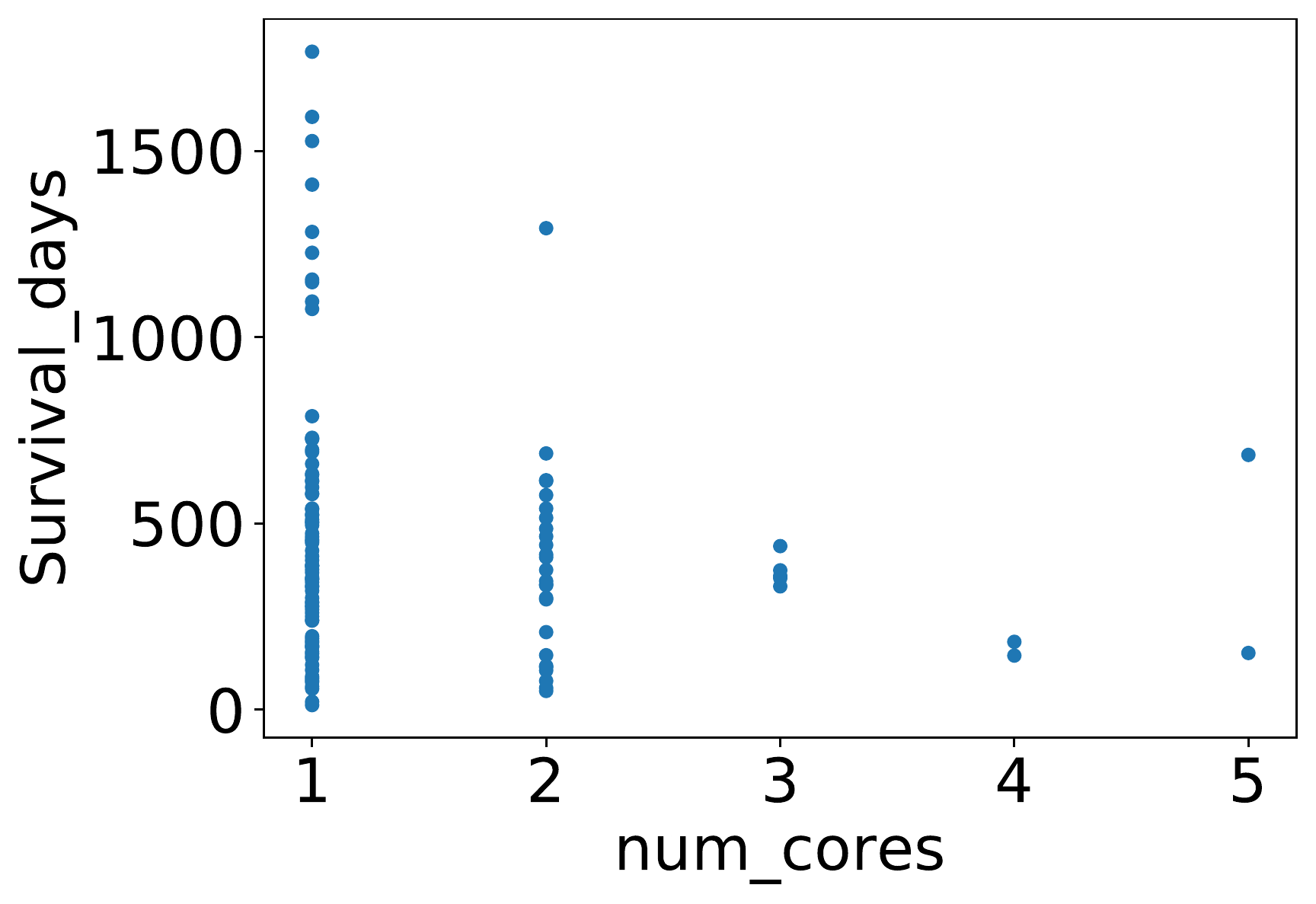}

                    \caption{Scatter plots of overall survival (in days) versus age, number of tumor components, and number of tumor cores, on the BraTS 2020 training dataset}
                    \label{fig:survival}
\end{figure}

We worked on the basis that any survival prediction will be approximate, and that the number of training cases is relatively small, from heterogeneous sources, and rather degraded owing to registration artifacts. We therefore picked two segmentation-derived features of the image which, based on the performance of our algorithm in segmentation, should be robust between centers:  number of disconnected tumor core regions (abbreviated below to 'number of cores')  and number of disconnected whole tumor regions (abbreviated below to 'number of tumors'), as segmented by our model.  As can be seen in Figure~\ref{fig:survival}, these measures are each predictive of survival on the training dataset, with increasing age, number of tumor components and number of tumor cores leading to reduced survival.

Initially, we built a least squares regression model predicting the survival time from age, number of disconnected cores  and number of tumors.  To avoid the survival prediction being biased by the longer survival times of especially long survivors, we replaced all survival times greater than 1000 days with 1000 (since these survival times are all far in excess of the 15 month cutoff relevant for the challenge).  Here we found that the image-derived features were largely unhelpful: ordinary least squares  models built on Age, number of tumor cores and number of tumors had coefficients for number of tumor cores and number of tumor compartments which were not significantly different from zero.

The actual validation for this challenge is based on classification accuracy (into the three survival classes) rather than regression performance.  The OLS regression models we built had an accuracy of 44 \%.  We subsequently trained a Random Forest model with a rather large number of trees (1000) with maximum depth 3 to predict the survival class, using Age alone, and also Age plus our image-derived features.  The random forest classifier on age alone also achieved 44 \% accuracy in cross-validation, while the model incorporating number of cores and number of tumors achieved  53 \% accuracy. This increased can be attributed to the tendency of the random forest model to predict more short-term survivors than the linear regression model. 

Our final model was a form of ensemble between the linear regression model and the classification  model, in which we allow the random forest classifier to override the prediction of the linear model.  To increase robustness, we only allow this when the RF is confident in its prediction: the predicted survival time was the output of the linear regression model, unless the classification model predicted (with probability 50 \% or more) a different class to the regression model.  In that case the predicted overall survival was moved to a fixed survival time corresponding to 10 months (in the case of low predicted survival) or 15 months (in the case of long predicted survival)

\section{Results and conclusions}

Our results on the BraTS testing set are presented in Tables~\ref{test_results_seg}, \ref{test_results_surv} and \ref{test_results_uncert}.  The mean Dice coefficients for our model were slightly improved over our 2019 method (the methodology for calculating Hausdorff distance changed from 2019 to 2020), which was ranked third in the 2019 challenge.  In terms of raw numeric challenge score, our method ranked joint 4th together with three other methods: however, no statistical significance was found between the score of the 4th ranked teams and the team ranked 3rd.    

Our accuracy of 0.617 in predicting patient survival ranked joint first in the survival challenge.  This is striking, as our method made use of only simple features, rather than a complex radiomic pipeline.  In particular, it is surprising that the number of tumor cores and number of tumor compartments has not been used by previous challenge participants though it may have been reflected in surrogate features defined by radiomic pipelines (for example, ratio of volume to surface area).  We believe our fusion of a prediction and classification model is also novel.

Our method for uncertainty estimation ranked first in the challenge, and was significantly better than the other methods entered in the challenge ($p<10^{-50}$).  Since this result arises from a challenge, it is difficult to analyse whether this dramatic difference arises from the superior uncertainty estimation of our model, or the observation that the challenge metric asymmetrically rewarded uncertainties inside and outside of the predicted volume.  Given the diminishing returns of comparing hard segmentations to manually derived ground truths, and the additional information that soft outputs characterizing uncertainty provide, this topic certainly merits further investigation.

\begin{table}
\centering
\begingroup
\setlength{\tabcolsep}{7pt} 
\renewcommand{\arraystretch}{1.5} 
\begin{tabular}{c|ccc|ccc}
 & \multicolumn{3}{c|}{Dice} & \multicolumn{3}{c}{Hausdorff95} \\
& ET      & WT     & TC     & ET        & WT       & TC        \\ \hline
Mean   & 0.82    & 0.90   & 0.83   & 13.6      & 4.7      & 22.0      \\
StdDev & 0.18    & 0.10   & 0.26   & 63.7      & 6.6      & 79.6     
\end{tabular} 
\endgroup
\vspace{1em}
\caption{Results on the BRATS 2020 test set (segmentation task), as delivered by the challenge organizers.}
\label{test_results_seg}
\end{table}

\begin{table}
\centering
\begingroup
\setlength{\tabcolsep}{7pt} 
\renewcommand{\arraystretch}{1.5} 
\begin{tabular}{c|ccccc}
 & Accuracy & MSE    & medianSE & stdSE   & SpearmanR \\ \hline
Value  & 0.617    & 391589 & 57600    & 1194915 & 0.378    
\end{tabular}
\endgroup

\vspace{1em}
\caption{Results on the BRATS 2020 test set (survival task), as delivered by the challenge organizers.}
\label{test_results_surv}
\end{table}

\begin{table}
\centering
\begingroup
\setlength{\tabcolsep}{7pt} 
\renewcommand{\arraystretch}{1.5} 
\begin{tabular}{c|ccc|ccc}
   & \multicolumn{3}{c|}{Dice AUC} & \multicolumn{3}{c}{Score} \\
  & WT       & TC       & ET      & WT      & TC      & ET     \\ \hline
Mean   & 0.91     & 0.85     & 0.84    & 0.97    & 0.95    & 0.95   \\
StdDev & 0.10     & 0.26     & 0.18    & 0.033   & 0.086   & 0.059 
\end{tabular}
\endgroup

\vspace{1em}
\caption{Results on the BRATS 2020 test set (uncertainty task), as delivered by the challenge organizers.}
\label{test_results_uncert}
\end{table}

\bigskip \noindent\textbf{Acknowledgements.}
This work was supported by the Swiss Personalized Health Network (SPHN, project number 2018DRI10).  Calculations were performed on UBELIX (\url{http://www.id.unibe.ch/hpc}), the HPC cluster at the University of Bern.
\bibliographystyle{splncs03}
\bibliography{MS}

\end{document}